\documentstyle[aps,12pt]{revtex}

\begin{document}

\def\m{{\bf m}}
\def\n{{\bf n}}
\def\r{{\bf r}}
\def\v{{\bf v}}
\def\x{{\bf x}}
\def\F{{\bf F}}
\def\R{{\bf R}}
\def\Z{{\bf Z}}
\def\del{\partial}
\def\Lap{\bigtriangleup}
\def\Div{{\rm div}\ }
\def\rot{{\rm rot}\ }
\def\curl{{\rm curl}\ }
\def\grad{{\rm grad}\ }
\def\Tr{{\rm Tr}}
\def\^{\wedge}
\def\real{{\rm re}}
\def\image{{\rm im}}
\def\goinf{\rightarrow\infty}
\def\goes{\rightarrow}
\def\bm{\boldmath}
\def\-{{-1}}
\def\inv{^{-1}}
\def\sqr{^{1/2}}
\def\isqr{^{-1/2}}

\def\eqn#1#2{\begin{equation} #1 \label{#2} \end{equation}}
\def\eqna#1{\begin{eqnarray} #1 \end{eqnarray}}
\def\reff#1{(\ref{#1})}
\def\vb#1{{\partial \over \partial #1}} 
\def\Del#1#2{{\partial #1 \over \partial #2}}
\def\Dell#1#2{{\partial^2 #1 \over \partial {#2}^2}}
\def\Dif#1#2{{d #1 \over d #2}}
\def\Lie#1{ {\cal L}_{#1} }
\def\diag#1{{\rm diag}(#1)}
\def\abs#1{\left | #1 \right |}
\def\rcp#1{{1\over #1}}
\def\paren#1{\left( #1 \right)}
\def\brace#1{\left\{ #1 \right\}}
\def\bra#1{\left[ #1 \right]}
\def\angl#1{\left\langle #1 \right\rangle}

\def\d{{\rm d}}
\def\xmy{x^2-y^2}
\def\xx{x^2-1}
\def\yy{1-y^2}
\def\px{\psi_x}
\def\py{\psi_y}
\def\pxx{\psi_{xx}}
\def\pyy{\psi_{yy}}
\def\pxy{\psi_{xy}}
\def\gx{\gamma_x}
\def\gy{\gamma_y}
\def\pr{\paren}
\def\xych{\mbox{$x\leftrightarrow y$}}
\def\exch{\leftrightarrow}
\def\Yp{{\rm Y}_p}
\def\bYp{\overline{\rm Y}_p}
\def\wa{&=&}
\def\wb{&\equiv &}

\def\br#1#2#3{\brace{#1#2#3}}
\def\f#1#2{f_{#1#2}}
\def\s#1{\sigma^{#1}}

\def\A{T}
\def\B{R}

\def\KY{Killing-Yano }

\def\abstract#1{\begin{center}{\large ABSTRACT}\end{center}
\par #1}
\def\title#1{\begin{center}{\large #1}\end{center}}
\def\author#1{\begin{center}{\sc #1}\end{center}}
\def\address#1{\begin{center}{\it #1}\end{center}}
\def\pubnum{253/COSMO-44}

\begin{titlepage}
\hfill
\parbox{6cm}{{TIT/HEP-\pubnum} \par April 1994 }
\par
\vspace{7mm}
\title{Static axisymmetric spacetimes with non-generic
world-line SUSY}
\vskip 1cm
\author{Masayuki TANIMOTO
\footnote{E-mail address: prince@phys.titech.ac.jp}}
\address{Department of Physics, Tokyo Institute of \\ Technology,
Oh-Okayama Meguroku, Tokyo 152, Japan}
\vskip 1 cm
\abstract{
The conditions for the existence of \KY tensors,
which are closely related to the appearance of non-generic
world-line SUSY,
are presented for static axisymmetric spacetimes.
Imposing the vacuum Einstein equation, the set of solutions
admitting \KY tensors is considered.
In particular, it is shown that
static, axisymmetric and
asymptotically flat vacuum solutions admitting \KY tensors
are only the Schwarzschild solution.
}

\vfill
PACS number(s): 4.20.Cv, 4.70.-s, 12.60.Jv
\end{titlepage}

\addtocounter{page}{1}

One of the most remarkable properties of the Kerr black hole is that,
in this background, particle motion is completely integrable.
{}From the point of view of canonical analysis,
this is a direct consequence of the existence of a non-trivial
Killing tensor $K_{\mu\nu}$ \cite{CarA,CarB,WP},
which gives rise to the associated constant of motion,
$K^{\mu\nu}p_\mu p_\nu$, quadratic in
the four-momentum $p_\mu$.
Namely, this constant of motion completes the maximal number of
constants of motion in conjunction with the other three
well-known constants of motion:
the energy coming from the time translation invariance
of the Kerr geometry, the angular momentum coming from
the axial symmetry of it, and the Hamiltonian.
More surprisingly, various field equations, particularly the
Dirac equation \cite{Cha}, separate in the Kerr geometry,
and \cite{CM} this fact is related to the existence of
the \KY tensor $\f\mu\nu$,
which is defined as an antisymmetric second rank tensor satisfying
the following Penrose-Floyd equation \cite{PF}:
\eqn{\brace{\mu\nu\lambda}\equiv2D_{(\mu}\f{\nu)}\lambda=0.}
{f1}
Here, $D_\mu$ represents the usual covariant derivative and,
for later convenience, we defined the braces.
Contracting repeated \KY tensors,
we obtain the Killing tensor associated with the \KY tensor:
$K_{\mu\nu}=f_\mu{}^\lambda\f\lambda\nu,\;
    D_{(\mu}K_{\nu\lambda)}=0$.
So, we might say that the \KY tensor is a square
root of the Killing tensor.

Recently, it has been shown by considering supersymmetric particle
mechanics that the \KY tensor can be understood as an object
belonging to a larger class of possible structures which generate
generalized supersymmetry algebras \cite{GRvH}.
This novel aspect has renewed people's interest in the \KY tensor
which has long been known to relativists.

It may be of some significance to ask conversely what kind of
background gravitational field admits the \KY tensor.
Are there any spacetime admitting a \KY tensor other than the Kerr
spacetimes?
While we already know that the existence of \KY tensor is directly
related to the appearance of
non-generic supersymmetry for the motion of spinning particle,
such an investigation will give another facet of
the structural properties of spacetimes.

A formal answer to this question has been already given
by Dietz and R\"{u}diger \cite{DR},
where all metrics admitting \KY tensors were presented
with no further assumptions restricting, say, the Ricci tensor.
However, interpretation of those metrics seems far from trivial.
In particular, the relation between the existence of \KY tensor
and isometries is not known.

In this paper, we take another approach to the same problem by
restricting from the beginning the metric to be static
and axisymmetric.
Using the prolate spheroidal coordinates $(x,y,\phi)$,
such metrics can be written in the form \cite{Pa}
\eqn{\d s^2=-e^{2\psi}\d t^2+\s2e^{-2\psi}
     \bra{e^{2\gamma}\pr\xmy\pr{\frac{\d x^2}{\xx}+
     \frac{\d y^2}{\yy}}+\pr\xx\pr\yy\d\phi^2},}
{prl1}
where
\eqn{\psi=\psi(x,y),\; \gamma=\gamma(x,y).}
{prl2}
Note that the parameter $\sigma$ can absorb possible
constants added to $\psi$ and $\gamma$, so that
the additional constants can always be taken zero
without loss of generality.
We will first describe, in terms of
$\psi(x,y)$ and $\gamma(x,y)$, the conditions
for the existence of \KY tensor.
Then we will see that the function
$\psi(x,y)$ which satisfies the vacuum Einstein equation
and admits a \KY tensor is determined
by its derivatives, $\px(p)$ and $\py(p)$,
at an arbitrary point $p=(x_0,y_0)$ (see \reff{second}).
Since at that time $\gamma(x,y)$ is determined
in terms of $\psi(x,y)$ through the Einstein equation,
the static axisymmetric vacuum solutions
admitting \KY tensors form a three dimensional set
parametrized by $\px(p)$, $\py(p)$ and $\sigma$.
However, when contracting this set by all possible diffeomorphisms,
we obtain, as the set of distinct solutions,
a complicated one dimensional set.
We then further restrict interest
to asymptotically flat cases, and finally
prove uniqueness of the Schwarzschild solution.

Before studying eq.\ \reff{f1} in terms of the metric
\reff{prl1}, we comment upon three properties of the metric
\reff{prl1}, which will be frequently used later.

First, we can immediately observe that
the metric \reff{prl1} is manifestly invariant
under the interchange \xych.
This implies, for example,
that if there is a correct expression for
$\psi$ and/or $\gamma$, the expression obtained
by interchanging \xych\ with $\px\exch\py$,
$\gx\exch\gy$, $\pxx\exch\pyy$, etc.
will also be a correct expression.
We call this interchange the ``conjugation''.

Second, the following formal transformation
\eqn{(t,\phi)\goes(i\sigma\phi,-i\sigma\inv t)}
{t-phi1}
preserves the characteristic form of the metric \reff{prl1},
i.e., the metric obtained after the transformation is still
designated by the (appropriately transformed) set of two functions
and one parameter, $(\psi(x,y),\gamma(x,y),\sigma)$.
In fact, this transformation is equivalent to the following
transformation for $\psi$ and $\gamma$:
\eqna{&&\hspace*{0em} (\psi,\gamma,\sigma)
               \goes(\psi,\gamma,\sigma)'=
               (\psi',\gamma',\sigma), \nonumber \\
      \psi'\wb -\psi+\rcp2\log\pr\xx\pr\yy, \nonumber \\
      \gamma'\wb \gamma-2\psi+\rcp2\log\pr\xx\pr\yy.
\label{t-phi2}
}
We call this transformation the $t$-$\phi$ transformation.
It is obvious from \reff{t-phi1} that if $(\psi,\gamma,\sigma)$ is a
solution of Einstein's equation,
then $(\psi',\gamma',\sigma)$ is also a solution,
which generally represents another spacetime.
This also implies that Einstein's equation is invariant
under the $t$-$\phi$ transformation \reff{t-phi2}.
Note that these transformations form a group isomorphic to
${\rm\bf Z}_2$,
i.e. $(\psi,\gamma,\sigma)''=(\psi,\gamma,\sigma)$.

Finally, there exists a 2-parameter diffeomorphism which also
preserves the characteristic form of \reff{prl1};
\eqna{&&\hspace*{9em} (x,y)\goes(x',y'), \nonumber \\
      x'\wb \rcp2\bigg(\sqrt{s^2(x^2+y^2)+2s(a+1)xy+(a+1)^2-s^2}
        \nonumber \\
        &&\hspace*{4em}
        +\sqrt{s^2(x^2+y^2)+2s(a-1)xy+(a-1)^2-s^2}\bigg),
        \nonumber \\
      y'\wb \rcp2\bigg(\sqrt{s^2(x^2+y^2)+2s(a+1)xy+(a+1)^2-s^2}
        \nonumber \\
        &&\hspace*{4em}-\sqrt{s^2(x^2+y^2)+2s(a-1)xy+(a-1)^2-s^2}
        \bigg),
\label{CPD1}
}
where $a$ and $s$ are arbitrary parameters.
Under this transformation,
the metric \reff{prl1} simply transforms as
\eqn{(\psi(x,y),\gamma(x,y),\sigma)
    \goes(\psi(x',y'),\gamma(x',y'),s\sigma).}
{CPD2}
We call these diffeomorphisms the characteristic-form-preserving
diffeomorphisms (CPDs)
\footnote{
The geometrical meaning of the CPDs becomes clear if one views them
in the cylindrical coordinates defined by
$(\rho,z,\phi)=(\sigma(x^2-1)^{1/2}(1-y^2)^{1/2},\sigma xy,\phi)$.
In these coordinates, the spatial part of the metric \reff{prl1}
takes the form
$e^{-2\psi}[e^{2\gamma}(\d\rho^2+\d z^2)+\rho^2\d\phi^2]$,
and the CPDs correspond to
the diffeomorphisms which map $(\rho,z)\goes(s\rho,sz+a)$.
Therefore the parameter $a$ is considered to represent
the translation along the axis of symmetry,
while $s$ represents the similarity transformation.}.

We are now in a position to study static axisymmetric spacetimes
which admit \KY tensors.
Our strategy is quite straightforward---
we simply write down all the components of the equation
$\br\mu\nu\lambda\equiv D_{(\mu}\f\nu{\lambda)}=0$ explicitly.
So, it would be worth noticing how many independent components
of eq.\ \reff{f1} exist.
For this purpose, note first that,
for the braces $\br\mu\nu\lambda$ defined in \reff{f1},
we can read the following symmetries;
\eqn{\br\mu\mu\mu=0
    \quad\mbox{\rm (all indices coincide)},}
{r3}
\eqn{\br\mu\mu\nu=-\br\nu\mu\mu=-\br\mu\nu\mu
    \quad\mbox{\rm (repeated indices exist)},}
{r2}
\eqn{\br\mu\nu\lambda=\br\nu\mu\lambda
    \quad\mbox{\rm (no repeated indices exist)}.}
{r1}
Now, it is easy to see that,
for the braces of the types both \reff{r2} and \reff{r1},
there are twelve independent components.
Thus, the total number of independent ones of
eq.\ \reff{f1} is 24, while that of $\f\mu\nu$ is six.

By explicit calculations of eq.\ \reff{f1},
we can see that the equations for components
$\br331$, $\br332$, $\br001$ and $\br002$
immediately imply that $\f12$ and $\f21$ must vanish,
and therefore the equations for components
$\br112$ and $\br221$ are identically satisfied.
Similarly, the equations for components
$\br013$, $\br310$, $\br023$, $\br320$, $\br301$, and $\br302$
demand that $\f30$ and $\f03$ vanish.

The remaining twelve components of eq.\ \reff{f1} are found
to be divided into two sets,
$\br113$, $\br312$, $\br223$, $\br123$, $\br231$ and $\br003$
which contain $\f13(=-\f31)$ and $\f23(=-\f32)$ only, and
$\br110$, $\br012$, $\br220$, $\br120$, $\br201$ and $\br330$
which contain $\f10(=-\f01)$ and $\f20(=-\f02)$ only.
For convenience we label these two sets ``\A'' and ``\B''
\footnote{As we shall see later, \A\ and \B\ are characterized by
``constraints'', which originate from, respectively,
the ``Time'' translation invariance and the ``Rotational''
invariance of our metric.},
respectively.
It is easy to check that \A\ and \B\ interchange under the $t$-$\phi$
transformation \reff{t-phi2}
with $\f13\exch\f10$ and $\f23\exch\f20$.
This means that, if we solve \A\ completely,
the solution of \B\ is automatically given.
For this reason, we shall mainly consider \A, since it is simpler
than \B.

Equation $\br003=2D_0\f03=0$ obviously contains
no derivatives of $\f\mu\nu$ due to the static nature
of the spacetime.
Its explicit form is
\eqn{\pr\xx\px\f13+\pr\yy\py\f23=0.}
{A0}
The other equations in \A\ contain derivatives
with respect to $x$ and/or $y$;
\def\:{=0\Leftrightarrow}

\noindent
$\br113\:$
\eqn{\f1{3,x}=\paren{\frac x\xmy-2\px+\gx}\f13+
    \frac\yy\xx\paren{\frac y\xmy+\py-\gy}\f23}
{A1}
$\br231\:$
\eqn{\f1{3,y}=-\paren{\frac{2y}\yy+\frac y\xmy+3\py-\gy}\f13-
    \paren{\frac x\xx-\frac x\xmy-\gx}\f23}
{A2}
$\br312\:$
\eqn{\f2{3,x}=\paren{\frac y\yy-\frac y\xmy+\gy}\f13+
    \paren{\frac{2x}\xx+\frac x\xmy-3\px+\gx}\f23}
{A3}
$\br223\:$
\eqn{\f2{3,y}=-\frac\xx\yy\paren{\frac x\xmy-\px+\gx}\f13-
    \paren{\frac y\xmy+2\py-\gy}\f23.}
{A4}
The remaining equation $\br123=0$ becomes trivial
if eqs.\ \reff{A2} and \reff{A3} are satisfied.
Note that \A\ is a set of simultaneous first order partial
differential equations with a ``constraint'' \reff{A0}.
\def\DeltaT{\Delta}
For consistency, the derivative of \reff{A0} with respect to
$x$ must vanish.
This yields
\eqn{\DeltaT\gy=
    \pr\xx\bra{\px\pxy-\py\pxx+\frac y\xmy\pr{\px^2-\py^2}}}
{Ad1}
with
\eqn{\DeltaT\equiv\bra{\pr\xx\px^2+\pr\yy\py^2}.}
{DeltaT}
The derivative of \reff{A0} with respect to $y$
yields the conjugation of \reff{Ad1}, which reads
\eqn{\DeltaT\gx=
    \pr\yy\bra{\py\pxy-\px\pyy+\frac x\xmy\pr{\px^2-\py^2}}.}
{Ad2}
We have also to impose the integrability condition for $\gamma$,
which gives rise to a non-trivial condition:
\eqn{\frac\del{\del x}
    \brace{(\mbox{the r.h.s. of \reff{Ad1}})/\DeltaT}=
    \frac\del{\del y}
    \brace{(\mbox{the r.h.s. of \reff{Ad2}})/\DeltaT}.
}
{integgamma}
Moreover, the integrability of $\f13$ must also be imposed.
Using \reff{Ad1} and \reff{Ad2} to eliminate derivatives
of $\gamma$, this implies
\eqna{\lefteqn{2\px^3\py
        +\pxx\py\bra{x\pr\xx\px^2-x\pr\yy\py^2-2y\pr\xx\px\py}}
            \nonumber \\
    && -\pxy\bra{x\pr\xx\px^3-3y\pr\xx\px^2\py}=
    \mbox{the conjugation of the l.h.s.}.
\label{Aint}}
Here, we have supposed $\px\neq0,\py\neq0$.
When imposing \reff{Aint},
we can see that the integrability of $\f23$ is trivially satisfied.

Now, we can summarize the results of the above calculations
as follows.
Once given a metric of the form \reff{prl1}, one can examine through
\reff{Ad1}$\sim$\reff{Aint} whether it admits non-trivial components
$\f13=-\f31$ and $\f23=-\f32$ of a \KY tensor.
Moreover, transforming the metric by \reff{t-phi2} and reexamining
\reff{Ad1}$\sim$\reff{Aint} for the new metric
\footnote{This operation is equivalent to transforming
eqs.\ \reff{Ad1}$\sim$\reff{Aint} by \reff{t-phi2} and
reexamining them for the original metric.},
one can also know
whether the original metric admits non-trivial components
$\f10=-\f01$ and $\f20=-\f02$.
As regards the other components of $\f\mu\nu$, they always vanish.

In order to get more definite consequences,
we here impose the vacuum Einstein equation given by
\eqn{\gx=\frac\yy\xmy
    \bra{x\pr\xx\px^2-x\pr\yy\py^2-2y\pr\xx\px\py},}
{gxein}
\eqn{\gy=\frac\xx\xmy
    \bra{y\pr\xx\px^2-y\pr\yy\py^2+2x\pr\yy\px\py},}
{gyein}
\eqn{0=2x\px+\pr\xx\pxx-2y\py+\pr\yy\pyy.}
{pein}
The integrability of $\gamma$ in the Einstein equation
is guaranteed by eq.\ \reff{pein}, so that we do not have to
consider another integrability condition \reff{integgamma}.
Moreover, by a direct calculation
eq.\ \reff{Aint} is found to be trivial if
eqs.\ \reff{Ad1}, \reff{Ad2} and \reff{gxein}$\sim$\reff{pein}
are satisfied.
As a result, it turns out that, for the present purpose,
we need only eqs.\ \reff{Ad1} and \reff{Ad2}
over the vacuum Einstein equation \reff{gxein}$\sim$\reff{pein}.

Eliminating $\gx$ and $\gy$ in \reff{Ad1} and \reff{Ad2} by
substituting \reff{gxein} and \reff{gyein},
we obtain three equations
for $\px$, $\py$, $\pxx$, $\pyy$, and $\pxy$,
and then we can algebraically solve these equations
for $\pxx$, $\pyy$, and $\pxy$ in terms of $\px$ and $\py$
(with $x$ and $y$);
\eqna{\pxx\wa F(\px,\py,x,y), \nonumber \\
  \pyy\wa F(\py,\px,y,x), \nonumber \\
  \pxy\wa G(\px,\py,x,y),
\label{second}
}
where $F$ and $G$ are functions with four arguments.
Although we do not write down these lengthy expressions
in explicit forms,
it is evident that they do always exist (except for $\px=\py=0$)
and are unique because the three equations are
linear in $\pxx$, $\pyy$, and $\pxy$.

Equation \reff{second} implies that, given the first order
derivatives $\px$ and $\py$ at an arbitrary point $p$,
higher derivatives of $\psi$ at $p$ are all given.
Thus, $\psi(x,y)$ and accordingly $\gamma(x,y)$ are
determined at least in the neighbourhood of $p$
in terms of two parameters $\px(p)$ and $\py(p)$.
This implies that the static axisymmetric vacuum solutions
admitting \KY tensors form a three dimensional set
parametrized by $\px(p)$, $\py(p)$, and $\sigma$,
which we denote as $\Yp$.
We must, however, recall the existence of
two dimensional diffeomorphisms (CPDs), which
contract $\Yp$ and yield a connected one dimensional set,
$\bYp$, as the set of distinct solutions.

Although we cannot find all solutions belonging to $\bYp$,
some features of $\bYp$ can be seen as follows.
First, one can confirm by direct substitutions into \reff{Ad1},
\reff{Ad2}, \reff{gxein} and \reff{gyein}
that the Schwarzschild solution
\eqn{\psi_{\rm Sch}=\rcp2\log\frac{x-1}{x+1}}
{sch}
certainly admits a \KY tensor.
We can further look for similar solutions in the form
\eqn{\psi=
    \log(x+1)^{\xi_1}(x-1)^{\xi_2}(1+y)^{\eta_1}(1-y)^{\eta_2},}
{IEe1}
where, $\xi_1$, $\xi_2$, $\eta_1$ and $\eta_2$ are parameters
which satisfies $\xi_1+\xi_2=\eta_1+\eta_2$ resulting
from the vacuum Einstein equation \reff{pein}.
For example, we find
\eqna{\psi\wa -\rcp2\log\pr\xx\pr\yy, \nonumber \\
  \psi\wa\rcp2\log\frac{(x-1)(1+y)}{(x+1)(1-y)},\quad \mbox{etc.}
\label{IEe2}
}
(and also their conjugations) as solutions admitting \KY tensors.
Now, note that eqs.\ \reff{sch} and \reff{IEe2} give rise to
representative metrics in $\Yp$ parametrized by $\sigma$
of which gauge orbits are generated by the CPDs.
One can investigate by the aid of computers what portions
of $\Yp$ the orbits share, and may find that the structure
of $\bYp$ is very complicated.
In particular, it seems that $\bYp$ has infinitely many branches.
However, we do not go into further details on this
digressive subject.

Our next step is to select out out of $\bYp$ all asymptotically flat
solutions admitting \KY tensors.
To this end, we expand $\psi$ at spatial infinity ($x\goinf$) as
\def\rf#1{{\rm f}_#1}
\eqn{\psi=-\frac{1}{x}+\frac{\rf3(y)}{x^3}+\frac{\rf4(y)}{x^4}
    +\cdots,}
{expand}
where the first term of the r.h.s. has been taken to coincide with
that of $\psi_{\rm Sch}$.
This can be always done by choosing the parameter $s$ in the
CPDs appropriately.
The second order term in the expansion, which corresponds to
the dipole component of the system,
has been taken zero by using the remaining
freedom of the CPDs parametrized by $a$.
Having restricted the form of the solutions to \reff{expand},
$\psi$ no longer has continuous freedom,
because the remaining 1-parameter freedom of the metric is
to be carried by the spatial conformal parameter $\sigma$.
Uniqueness of the expression \reff{second} suggests that the
Schwarzschild solution \reff{sch} is the unique solution of
the form \reff{expand}.
In fact, this is confirmed by directly solving the
expanded equations of \reff{second}.
This fact implies more precisely that the solutions of the form
\reff{expand} admitting non-trivial components
$\f13=-\f31$ and $\f23=-\f32$ of
\KY tensors are only the Schwarzschild solutions
parametrized by $\sigma$.

What we have to do next is to examine if there are asymptotically
flat solutions which
admit non-trivial components $\f10=-\f01$ and $\f20=-\f02$ of
\KY tensors.
However, it is found as follows that no such solutions exist.
Namely, if there were such an asymptotically flat solution,
the asymptotic form of the Schwarzschild solution
$\psi\goes-1/x$
should satisfy the asymptotic equations of the $t$-$\phi$ transformed
version of equations \reff{second}.
However, this is not the case, and we conclude that the claim
is true.

Gathering all results, we have proved that,
{\it modulo diffeomorphisms,
static, axisymmetric,
and asymptotically flat vacuum solutions admitting
\KY tensors are only the Schwarzschild solutions.}

In summary, we first supposed spacetimes to be static and
axisymmetric, then finally derived
the Schwarzschild solutions as the unique
asymptotically flat vacuum solutions admitting \KY tensors.
Note that, as shown in \cite{DR},
in order to admit a \KY tensor a spacetime
must be of Petrov type D (or IV).
Although for Petrov type D spacetimes
all vacuum solutions are known \cite{Ki},
many of them are difficult to interpret.
In particular,
it is not trivial how many spacetimes which have isometries exist
and which are the asymptotically flat spacetimes.
Our approach has, contrary to this, an advantage that
physical interpretation of the results are manifest.

If we consider the stationary cases instead of static ones,
a new function $\omega(x,y)$ must be introduced in \reff{prl1}
\cite{Pa};
$$\d s^2=-e^{2\psi}(\d t-\omega\d\phi)^2+\s2e^{-2\psi}
     \bra{e^{2\gamma}\!\pr\xmy\!\pr{\frac{\d x^2}{\xx}
        +\frac{\d y^2}{\yy}}+\pr\xx\!\pr\yy\d\phi^2}.$$
The function $\omega$ glues the sets \A\ and \B\ in a so
complicated way that we have not tried such cases here.
Nevertheless, we believe that the extra conditions for the
existence of \KY tensors would still be so restrictive that,
as asymptotically flat vacuum solutions,
only the Kerr solutions would admit \KY tensors.

\bigskip

I wish to thank professor A. Hosoya for bringing \cite{GRvH}
to my attention and for continuous encouragement.
I would also like to thank K. Watanabe for useful suggestions
and T. Koike for discussions.

\vfill


\begin{references}

\bibitem{CarA} B. Carter, Commun. Math. Phys. 10 (1968) 280.

\bibitem{CarB} B. Carter, Phys. Rev. 174 (1968) 1559.

\bibitem{WP} M. Walker and R. Penrose, Commun. Math. Phys.
    18 (1970) 265.

\bibitem{Cha} S. Chandrasekhar, Proc. R. Soc. Lond. A349 (1976) 571

\bibitem{CM} B. Carter and R.G. McLenaghan, Phys. Rev. D19 (1979)
    1979.

\bibitem{PF} R. Penrose, Ann. NY Acad. Sci. 224 (1973) 125;
    R. Floyd, PhD. Thesis, London (1973).

\bibitem{GRvH} G.W. Gibbons, R.H. Rietdijk and J.W. van Holten,
    Nucl. Phys. B404 (1993) 42.

\bibitem{DR} W. Dietz and R. R\"{u}diger, Proc. R. Soc. Lond.
    A375 (1981) 361; Proc. R. Soc. Lond. A381 (1982) 315.

\bibitem{Pa} A. Papapetrou, Ann. Physik 12 (1953) 309.

\bibitem{Ki} W. Kinnersley, J. Math. Phys. 10 (1969) 1195.

\end{references}
\end{document}